\documentclass[jcp,showpacs,twocolumn,superscriptaddress,amsmath,amssymb,longbibliography]{revtex4-1}

\usepackage{graphicx}
\usepackage{dcolumn}
\usepackage{bm}
\usepackage{dcolumn,amsmath}
\usepackage[normalem]{ulem}

\usepackage[utf8]{inputenc}
\usepackage[T1]{fontenc}
\usepackage{mathptmx}
\usepackage{etoolbox}
\usepackage{color}

\draft 

\begin{document}


\title{Full-dimensional quantum scattering calculations of rovibrationally excited HD+HD collisions} 


	
\author{Bikramaditya Mandal}
\affiliation{Department of Chemistry and Biochemistry, University of Nevada, Las Vegas, Nevada 89154, USA}

\author{Hubert J\'o\'zwiak}
\affiliation{Institute for Molecules and Materials, Radboud University, Nijmegen, The Netherlands}
\affiliation{Institute of Physics, Faculty of Physics, Astronomy and Informatics, Nicolaus Copernicus University in Toru\'n, Grudzi\k{a}dzka 5, 87-100 Toru\'n, Poland}

\author{Piotr Wcis{\l}o}
\affiliation{Institute of Physics, Faculty of Physics, Astronomy and Informatics, Nicolaus Copernicus University in Toru\'n, Grudzi\k{a}dzka 5, 87-100 Toru\'n, Poland}


\author{Naduvalath Balakrishnan}
\thanks{Author to whom correspondence should be addressed: \\ naduvala@unlv.nevada.edu}
\affiliation{Department of Chemistry and Biochemistry, University of Nevada, Las Vegas, Nevada 89154, USA}


\date{\today}

\begin{abstract}
Full-dimensional quantum scattering calculations are reported for ro-vibrational transitions in HD+HD collisions using a highly accurate interaction potential for the H$_2$-H$_2$ system. Several near-resonant ro-vibrational transitions are identified that conserve the overall rotational angular momentum  and nearly conserve the internal energy of the collision partners. Key anisotropic terms that drive the rotational transitions and angular momentum partial waves that contribute to low energy resonant features in the energy dependence of the cross sections are identified. The computed results are in agreement with total cross sections reported in previous experimental results, including resonant features in the energy dependence of the cross section. In particular, low-energy cross sections show a strong resonant feature associated with an $l=3$ partial wave in the incident channel. Rate coefficients for several inelastic rotational and ro-vibrational transitions are reported for temperatures ranging from $0.1$ K to $200$ K and they display a maximum between $1$ K-$10$ K reflecting the important contributions from the $l=3$ shape resonance that occurs around 2.5 K.

\end{abstract}

\pacs{}

\maketitle 

\section{Introduction} 
Being the most abundant molecule in the universe, collisions of H$_2$ with itself and other molecules are of key interest in many astrophysical environments~\cite{Stancil1998, Wan2018}. Inelastic collisions and reactions involving small molecules, such as H$_{2}$ and its deuterated counterpart HD, play a crucial role in the chemistry of the early universe~\cite{Stancil1998, Glover2008, Galli1998, Wan2019,Balakrishnan2018, Puy1993, Galli2002}, cold interstellar media (ISM)\cite{Wan2019, Lacour2005, Liszt2015}, star forming regions~\cite{McGreer2008, Hirano2015, Ripamonti2007}, and planetary atmospheres.  Unlike H$_{2}$ which has  no permanent dipole moment, the detection of HD  is easier, thanks to its small but finite dipole moment. Spectral line corresponding to the  $j=1 \to j'=0$ rotational transition in HD has been detected by the Herschel space observatory~\cite{Joblin2018, Bergin2013} and the Infrared Space Observatory (ISO)  using the long wavelength spectrometer~\cite{Wright1999, Polehampton2002}. Rotational line arising from the  $j=4 \to j'=3$ transition in HD has also been observed by the Atacama large millimeter array (ALMA) and the Spitzer space telescope~\cite{Kamaya2003, Wright1999, Neufeld2006}. These observations have spurred much interest in collisions of rotationally excited HD with its dominant collision partners, H$_2$, He and CO. Although its abundance in the interstellar medium is relatively small compared to H$_{2}$ with a D/H ratio of $\sim$ $3\times 10^{-5}$, HD is thought to play a key role in the cooling of the primordial gas due to its small dipole moment~\cite{Stancil1998, Wan2019, Wan2018, Flower2007, Flower2000}.

Collisions of HD with H$_2$, D$_2$, and He have also been the topic of several recent experimental studies by Zare and coworkers employing co-propagation of the two collision partners in the same molecular beam~\cite{Perreault2016, Perreault2017, Perreault2018, Perreault2018NatChem, Perreault2019, Perreault2022JPCL}. This allows small relative velocities for the collisions which limit the number of angular momentum partial waves to a few, mostly $l=0-3$. The co-expansion combined with the Stark-induced adiabatic Raman passage (SARP) technique allows control of the HD bond axis alignment relative to the initial velocity vector. The alignment is controlled by selecting specific $m_j$ states (magnetic quantum number) of the initial molecular rotational level $j$ or a superposition of $m_j$ states. For example, for $j=2$, $m_j=0$ corresponds to a horizontal alignment while a superposition of $m_j=0,\pm 2$ corresponds to a vertical alignment. The experiments, measuring angular distribution of the scattered HD molecule in a rotationally inelastic collision, provide sensitive probe of the interaction potential for these benchmark systems. Several theoretical studies have provided explicit simulation of these experiments, including specific angular momentum partial wave(s) that imprint distinct signatures in the measured angular distribution~\cite{Croft2018, Croft2019, Morita2020word64, Morita2020word65, Jambrina2019, Jambrina2021, mandal2024HDD2, Mandal_HeHD_2025, sanz2025cold, perez-hernando2025resonance, Croft2023}. 

In the course of these experiments that probed $\Delta j=-1$ \& $\Delta j=-2$ transitions in HD($v=1,j=2$)+H$_2$/D$_2$~\cite{Perreault2017,Perreault2018,Perreault2018NatChem,Perreault2019,Perreault2022JPCL} and $\Delta j=-2$ transition in D$_2$($v=2,j=2$)+D$_2$($v=2,j=2$)~\cite{Zhou2022} collisions, a new potential energy surface (PES) for H$_2$-H$_2$ collisions that allows the study of collisions between highly vibrationally excited H$_2$ and HD molecules have been reported~\cite{Zuo2021}. While possibility of studying four-center reactions such as HD($v\geq 4$)+HD($v\geq4$)$\to$ H$_2$+D$_2$ using the SARP techniques is exciting, the experiments have not been performed yet. However, recent theoretical studies have explored this process using quasi classical trajectory (QCT) methods~\cite{Liu_H2H2_2024}. 

In earlier full-dimensional quantum calculations of H$_2$+H$_2$ collisions, the importance of near-resonant rotational transitions that conserve the total molecular rotational angular momentum and nearly conserve the internal energy have been discussed~\cite{Quemener_H2H2_2008, Bala_JCP_H2H2_2011, santos.balakrishnan.ea:quantum, Samantha_H2H2_2013}. For example, it was shown that in collisions of H$_2$($v=1,j=0$)+H$_2$($v=0,j=2$) the most dominant inelastic channel is H$_2$($v=1,j=2$)+H$_2$($v=0,j=0$)~\cite{Bala_JCP_H2H2_2011}. This near-resonant transition that conserves the total internal rotational angular momentum of the two molecules has an energy defect of about 24.45 K and has a cross section that is two orders of magnitude larger than the pure rotational quenching of the H$_2$($v=0,j=2$) collision partner. The possibility of such quasi-resonant transitions in HD+HD collisions has not been explored so far though the energy gap for such transitions in HD is smaller due to the possibility of $\Delta j=\pm 1$ transitions.

Due to its importance as a benchmark system for collisional studies involving 4-atom systems as well as astrophysical interest, the H$_4$ system has been the topic of many electronic structure calculations over the last few decades. This includes both four-dimensional (4D) PESs within a rigid rotor approximation for the H$_2$ molecules and full six-dimensional (6D) PESs. The 4D PESs of Diep and Johnson~\cite{Diep_JCP_2000,Diep_JCP_Erratum_2000} and that of Patkowski et al.~\cite{Patkowski_H4_2008} have been used in  prior calculations of rotational energy transfer in H$_2$+H$_2$~\cite{Lee_2008,Wan2018} and H$_2$+HD collisions~\cite{Wan2019}. Flower~\cite{Flower_H4_1998, Flower_H4_2000, Flower_1999} and Flower and Rueff~\cite{Flower_Rueff_H4_1998, Flower_Rueff_H4_1999, Flower_Rueff_1999} have also reported extensive calculations of rotational energy transfer in H$_2$+H$_2$ and H$_2$+HD collisions using a PES developed by Schwenke~\cite{Schwenke_H4_1988}. Full dimensional PESs have been reported by Boothroyd et al.~\cite{Boothroyd_JCP_2002}, Hinde~\cite{Hinde2008}, Garberoglio et al.~\cite{H4_PES_Garberoglio}, and Zuo et al.~\cite{Zuo2021}. Qu\'em\'ener et al.~\cite{Goulven2008, Goulven2009}, Balakrishnan et al.~\cite{Bala_JCP_H2H2_2011, Balakrishnan2018},  and dos Santos et al.~\cite{santos.balakrishnan.ea:quantum, Samantha_H2H2_2013} have reported extensive calculations of rovibrational transitions in H$_2$+H$_2$ collisions using the potential energy surfaces of Boothroyd et al.~\cite{Boothroyd_JCP_2002} and Hinde~\cite{Hinde2008}. Very recently, a new 6D PES for the H$_4$ system by Jankowski, Patkowski, and Szalewicz (hereafter referred to as the JPS surface) has been adopted by J{\'o}{\'z}wiak et al.~\cite{jozwiak2024accurate} to compute beyond-Voigt line-shape parameters for H$_{2}$-perturbed $[R(0)-R(2)]$ rotational lines in HD. The form of the long-range part of this PES was taken from the 4D PES of Patkowski et al.~\cite{Patkowski_H4_2008}. Due to its larger basis set and higher level of theory used in the electronic structure calculations, this PES can be considered the most accurate full-dimensional surface currently available for modeling inelastic scattering between two H$_2$ molecules and their isotopomers. We adopt this surface for the computations reported in this work. The computations were carried out using a modified version of the TwoBC~\cite{Krems2006} quantum scattering code.

We would also like to emphasize that while H$_2$+H$_2$ and H$_2$+HD collisions have been studied quite extensively due to their significance in astrophysics, HD+HD collisions have received much less attention. Besides an experimental study reported by Johnson et al.~\cite{johnson1979total} in which energy resolved total cross sections for HD+HD collisions were measured, the only recent quantum mechanical calculations of the state-to-state cross sections that we are aware of are by Sultanov et al.~\cite{sultanov2009state, sultanov2012ultracold} within the 4D rigid rotor formalism. Quantum scattering calculations of  beyond-Voigt line-shape parameters for  self-perturbed rovibrational transition in HD have also been recently reported by Cygan et al.~\cite{cygan2025dispersive} using the JPS PES averaged over the vibrational wave functions of HD. Full rovibrational quantum calculations of HD+HD collisions have not been reported so far. Classical trajectory and semiclassical calculations have  previously been reported, respectively, by Gelb and Alper~\cite{Gelb1979}, Brown and Longuemare~\cite{brown1990calculation}, and Cacciatore and Billing~\cite{Billing1992}.

The paper is organized as follows: Section \ref{sec:methods} provides a brief description of the methodology, including details of the PES and the scattering calculations. Results are presented in section \ref{sec:results} and a summary of our findings is given in section \ref{sec:summary}.
	
\section{Methods}
\label{sec:methods}
\subsection{Potential Energy Surface}
Within the Born-Oppenheimer (BO) approximation, the interaction potential for HD+HD collisions is the same as that of H$_2$+H$_2$, except for the minor diagonal BO correction (DBOC) which is  mass dependent. This correction term is included in the JPS PES, but its magnitude is negligibly small compared to the uncertainty of the PES and does not affect the outcome of quantum scattering calculations. Further, isotope-dependent interaction potentials for He+HD and He+D$_2$ collisions with appropriate mass-dependent DBOC corrections for HD and D$_2$  yielded nearly identical results as that derived from their He+H$_2$ counterpart~\cite{Mandal_HeHD_2025}. For the scattering calculations, the angular dependence of the interaction potential is expanded in bispherical harmonics~\cite{Goulven2009}. The potential for HD-HD interaction is computed by shifting the center-of-mass (COM) of the two H$_{2}$ molecules to that of HD and recomputing the potential in Jacobi coordinates appropriate for HD+HD interaction:
%
\begin{equation}\label{eqn:exp_projection}
	V(\vec{r_{1}}, \vec{r_{2}}, \vec{R})= \sum_{\lambda_{1}, \lambda_{2}, \lambda_{12}} C_{\lambda_{1}, \lambda_{2}, \lambda_{12}}(r_{1}, r_{2}, R) Y_{\lambda_{1}, \lambda_{2}, \lambda_{12}}(\hat{r_{1}}, \hat{r_{2}}, \hat{R})
\end{equation}
where,
\begin{equation}\label{eqn:y_function}
	\begin{split}
		Y_{\lambda_{1}, \lambda_{2}, \lambda_{12}}(\hat{r_{1}}, \hat{r_{2}}, \hat{R})&= \sum_{m_{1}, m_{2}, m_{12}} \langle \lambda_{1} m_{1} \lambda_{2} m_{2}|\lambda_{12} m_{12}\rangle\\
		&~\times Y_{\lambda_{1}, m_{1}}(\hat{r_{1}})Y_{\lambda_{2}, m_{2}}(\hat{r_{2}}) Y^{*}_{\lambda_{12}, m_{12}}(\hat{R}).
	\end{split}	
\end{equation}
Unlike H$_2$+H$_2$ collisions, which involve only even-order terms in the expansion, due to the shift in the COM of HD relative to that of H$_2$, both even and odd  terms are present in the expansion coefficients

\begin{figure}
\centering
\includegraphics[width=0.5\textwidth, keepaspectratio,]{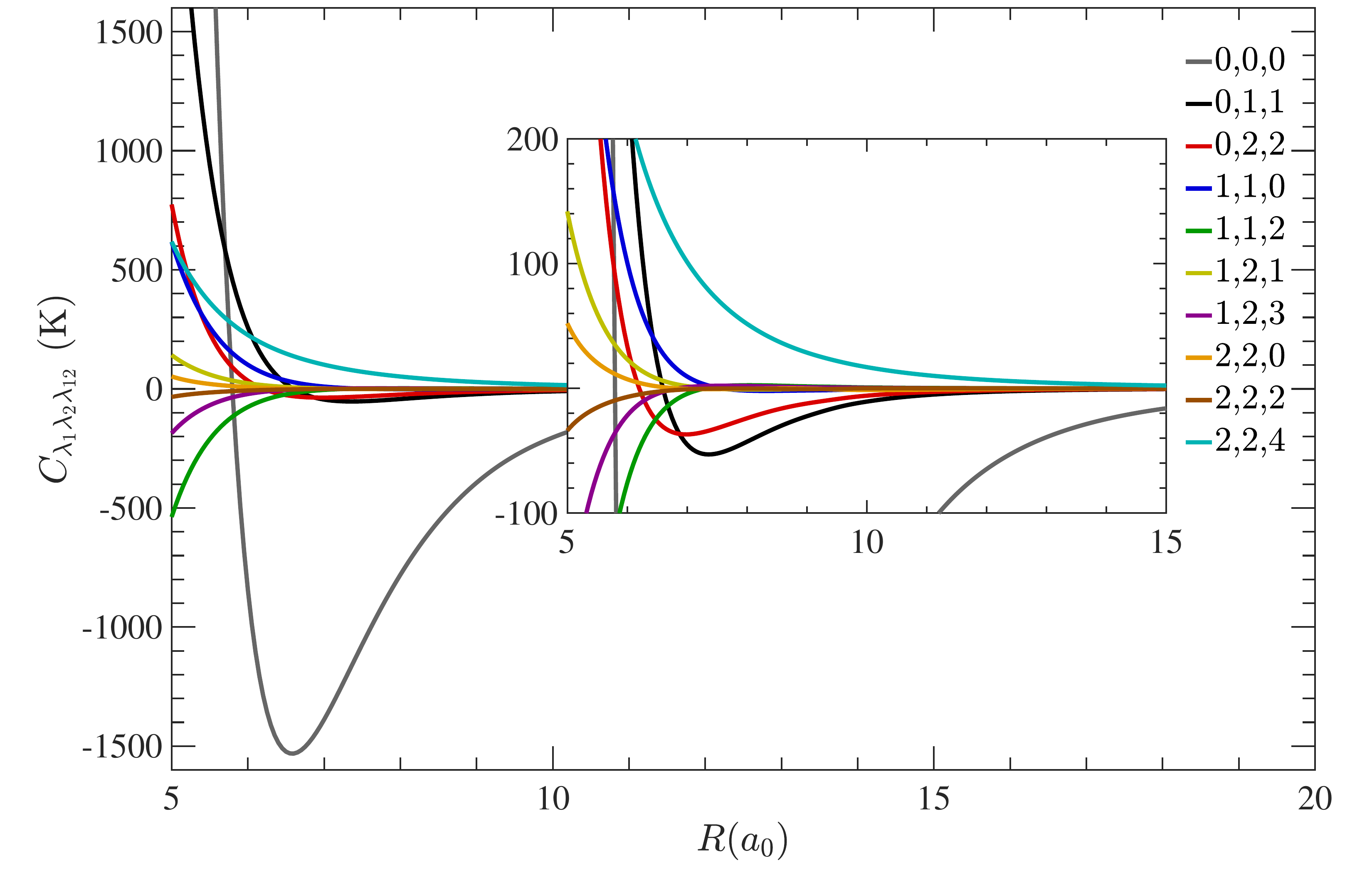}
\caption{The leading expansion coefficients of the HD+HD interaction  as  functions of the intermolecular separation. Different terms are labeled by $\lambda_{1}, \lambda_{2}$, and $\lambda_{12}$ in that order. The inset provides an enlarged view of the leading  terms in the region of the van der Waals potential well. Note that the bond lengths of both the HD molecules were fixed at 1.44 $a_{0}$.}
\label{fig:Fig_exp_coeffs}
\end{figure}

Figure~\ref{fig:Fig_exp_coeffs} displays the leading expansion coefficients as obtained  from eqs. (\ref{eqn:exp_projection}-\ref{eqn:y_function}) as a function of the COM separation between the two HD molecules keeping bond lengths of both the HD molecules fixed at 1.44 $a_{0}$. The inset in Fig.~\ref{fig:Fig_exp_coeffs} provides an enlarged view of the expansion coefficients in the vicinity of the van der Waals potential well. The leading anisotropic term is $(\lambda_{1}, \lambda_{2}, \lambda_{12}) =(2, 2, 4)$, which drives the $|\Delta j|=2$ transitions. The remaining anisotropic terms, $(0, 1, 1)$, $(1, 1, 0)$, and $(1, 1, 2)$ that drive the $|\Delta j|=1$ transitions  are relatively small in magnitude. Based on the magnitude of the anisotropic terms, with no consideration of the energy gap, one  can expect  that $|\Delta j|=2$ transitions are likely to be more intense than $|\Delta j|=1$ transitions.
For the scattering calculations, it was found that all terms upto $\lambda_{1}=\lambda_{2} \le 6$ for HD were sufficient to yield converged results on the JPS PES. Upon projecting expansion coefficients on the basis functions, that include monomer vibrations, we rigorously include all couplings (including their vibrational/rotational dependence).

\subsection{Scattering Calculations}
The quantum scattering calculations were carried out using a modified version of the TwoBC code~\cite{Krems2006} that implements  full-dimensional coupled-channel (CC) calculations of collision between two ($^1\Sigma$) vibrating molecules in the total angular momentum representation of Arthurs and Dalgarno~\cite{arthurs1960theory}. The method has been applied to prior studies of H$_{2}$-H$_{2}$  and its isotopologues~\cite{Balakrishnan2018, Goulven2008, Goulven2009,Bala_JCP_H2H2_2011,santos.balakrishnan.ea:quantum,mandal2024HDD2}. Therefore, we  provide only a brief outline of the scattering formalism to introduce pertinent quantum numbers  and notations.

The solution of the time-independent Schr\"{o}dinger equation within the CC method yields the scattering matrix, \textbf{S}, from which state-to-state cross sections for ro-vibrational transitions are extracted. We assign  variables $n$ and $n'$ to label the initial and final combined molecular states (CMSs) of the collision partners. Specifically, $n \equiv v_{1} j_{1} v_{2} j_{2}$  and $n' \equiv v_{1}' j_{1}' v_{2}' j_{2}'$ for the pair of HD molecules before and after the collision. The integral cross section (ICS) for state-to-state rovibrational transitions for identical particle scattering at a given collisional energy ($E_{c}$) is given by

\begin{equation}\label{eqn:ICS_eqn}
	\begin{split}
		\sigma_{n \to n'}(E_{\rm{c}})= & \frac{\pi(1+\delta_{v_{1}v_{2}}\delta_{j_{1}j_{2}})(1+\delta_{v'_{1}v'_{2}}\delta_{j'_{1}j'_{2}})}{k_{n}^{2}(2j_{1}+1)(2j_{2}+1)} \times \\
		& \sum_{J, j_{12}, j_{12}', l, l'} (2J+1)|T^{J}_{n,l,j_{12},n',l',j_{12}'}|^{2},
	\end{split}	
\end{equation}
where $k_{n}^{2}=2\mu E_{\rm{c}}/\hbar^{2}$, $E_{\rm{c}}=E-E_{n}$, $E_{n}$ is the asymptotic energy of channel $n$, $E$ is the total energy, $\mu$ is the reduced mass of the two HD molecules, and $T^{J}=1-S^{J}$. The  total angular momentum of the four-atom system is given by $\bf{J=l+j_{12}}$, where $\bf{l}$ is the orbital angular momentum and $\bf{j_{12}=j_{1}+j_{2}}$ is the total molecular rotational angular momentum with $J$, $l$, and $j_{12}$ denoting the corresponding quantum numbers.

In the helicity representation, the scattering amplitudes as a function of the scattering angle $\theta$ is given by~\cite{Schaefer1979}
\begin{equation}\label{eqn:scattering_amplitude}
	\begin{split}
		q_{n,m \to n', m'} (\theta)= & \frac{1}{2k_{n}} \sum_{J} (2J+1) \\
		& \sum_{j_{12}, j_{12}', l, l'} i^{l-l'+1} T^{J}_{n,l,j_{12},n',l',j_{12}'} d^{J}_{m_{12},m_{12}'}(\theta) \\
		& \times \langle j_{12} m_{12} J -m_{12}|l 0 \rangle \langle j_{12}' m_{12}' J -m_{12}'|l^{'} 0 \rangle \\
		& \times \langle j_{1} m_{1} j_{2} m_{2}|j_{12} m_{12} \rangle \langle j_{1}' m_{1}' j_{2}' m_{2}'|j_{12}' m_{12}' \rangle,
	\end{split}	
\end{equation}
where $d^{J}_{m_{12},m_{12}'}(\theta)$ is an element of the Wigner reduced rotation matrix, $m \equiv m_{1}, m_{2}, m_{12}, ~m'\equiv m_{1}', m_{2}', m_{12}'$, and the quantities in angular brackets $\langle ....|..\rangle$ are Clebsch-Gordan coefficients. For isotropic collisions, the differential cross sections (DCSs) for rovibrational state-resolved transitions are obtained by summing over all final $m'$-states and averaging over initial $m$-states:

\begin{equation}\label{eqn:dcs_eqn}
	\frac{d\sigma_{n \to n'}}{d\Omega}= \frac{(1+\delta_{v_{1}v_{2}}\delta_{j_{1}j_{2}})(1+\delta_{v'_{1}v'_{2}}\delta_{j'_{1}j'_{2}})}{(2j_{1}+1)(2j_{2}+1)} \sum_{m,m'} \left\vert q_{n,m \to n',m'} \right\vert ^{2},
\end{equation}
where the solid angle $d\Omega=$sin$\theta d\theta d\phi$ and $\phi$ is the azimuthal angle.

A basis set consisting of three vibrational levels, $v=0-2$ of the two HD molecules with  five rotational states $j=0-4$, in each of the vibrational levels were adopted for the calculations. This led to nearly 120 CMSs considering identical collision partners. Computations were performed for kinetic collision energies ranging from  $E_{c}=10^{-3}$ K $-10^{3}$ K and  $J=0-40$. For collision energies below 10 K only $J=0-10$ were required to yield converged ICS. The total number of channels for $J=40$ and $E_{c}=10^{3}$ K is 1500. The CC equations were integrated from $R=3$ to $R=103~ a_{0}$ with a step size of $\Delta R=0.05~ a_{0}$. Results are converged to within 1\% with respect to the basis set and other parameters involved in the numerical solution of the Schr\"{o}dinger equation.

Rate coefficients for state-to-state ro-vibrational transitions at a given temperature $T$ are given by averaging the corresponding cross sections over a Maxwell-Boltzman distribution of collision energies
\begin{equation}
	\begin{aligned}
		k_{n\to n'}(T)&=\frac{v_{\rm{ave}}(T)}{(k_{\rm{B}}T)^{2}}\times\\
		&\int\limits_{E_{c}=0}^{\infty}E_{c} ~\sigma_{n\to n'}(E_{c})~e^{-\frac{E_{c}}{k_{\rm{B}}T}}dE_{c},
	\end{aligned}
	\label{maxwell_blotzmann_eqn}
\end{equation}
where $k_{\rm{B}}$ is the Boltzmann constant and  the average collision velocity $v_{\rm{ave}}(T)=\sqrt{8k_{\rm{B}}T/\pi\mu}$.
	
\section{Results}
\label{sec:results}
\begin{figure}
	\centering
	\includegraphics[width=0.5\textwidth, keepaspectratio,]{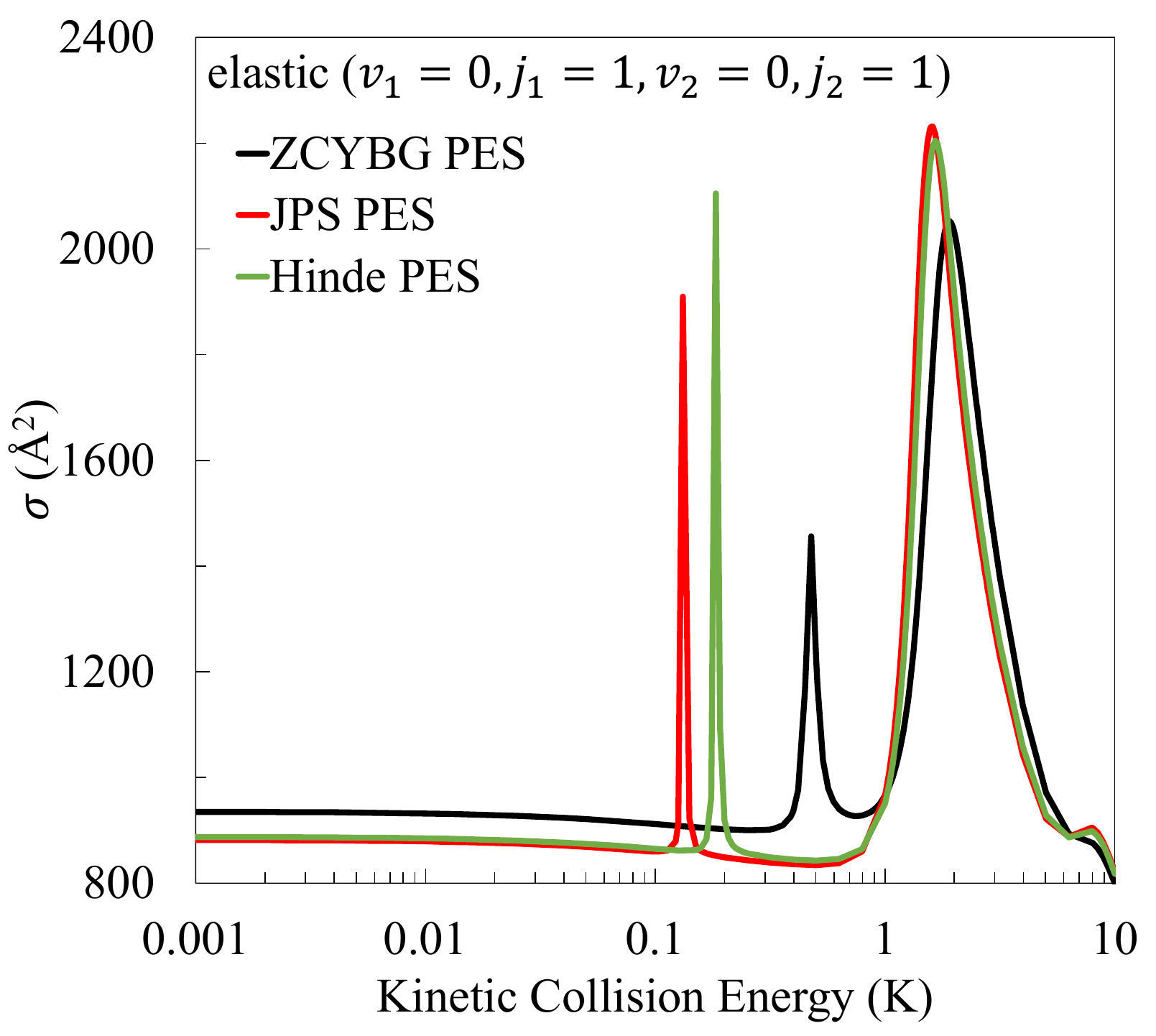}
	\caption{Elastic cross sections for ortho-H$_2$+ortho-H$_2$ collisions as functions of the collision energy computed using the ZCYBG, JPS, and Hinde PESs for the initial state $v_{1},j_{1},v_{2},j_{2}=0,1,0,1$.}
	\label{fig:H4_ini_0101}
\end{figure}

As discussed in the introduction, several 6D PESs are available for probing collisions between two H$_2$ molecules or their isotopologues. It would be instructive to compare cross sections for elastic H$_2$+H$_2$ collisions at low energies on some of the newer PESs for the H$_4$ system. We used the PESs of Hinde~\cite{Hinde2008}, Zuo et al.~\cite{Zuo2021} (hereafter referred to as ZCYBG) and JPS~\cite{jozwiak2024accurate} for this purpose. The resulting elastic cross sections for  $v_{1},j_{1},v_{2},j_{2}=0,1,0,1$ initial state as functions of the collision energy for $E_c=10^{-3}$ K$-10$ K are shown in Figure~\ref{fig:H4_ini_0101}. All three PESs yield nearly identical results for $E_c>1$ K, including a shape resonance located at $E_c\approx 2$ K. A sharp resonance feature is observed below 1 K but its position is significantly different for the three PESs with the ZCYBG PES predicting it at $\sim$ 0.5 K while the Hinde and JPS  surfaces predicting it between 0.1 and 0.2 K. Overall, it appears that the results on the Hinde and JPS PESs are in close agreement. Note that these two PESs are computed with the coupled-cluster singles, doubles and perturbative triples (CCSD(T)) method and they include corrections beyond the  CCSD(T) level of theory.  However, the JPS PES is computed with a larger basis set and it also accounts for the DBOC term but for the H$_4$ system as discussed earlier. The ZCYBG PES is computed using the multi-reference configuration interaction (MRCI) method and it is designed to describe collisions of highly vibrationally excited H$_2$ molecules. Based on these comparisons and earlier discussions of the PESs, we adopt the JPS PES for the HD+HD computations reported here.

\subsection{Comparisons with prior experimental results}
\begin{figure}
	\centering
	\includegraphics[width=0.5\textwidth, keepaspectratio,]{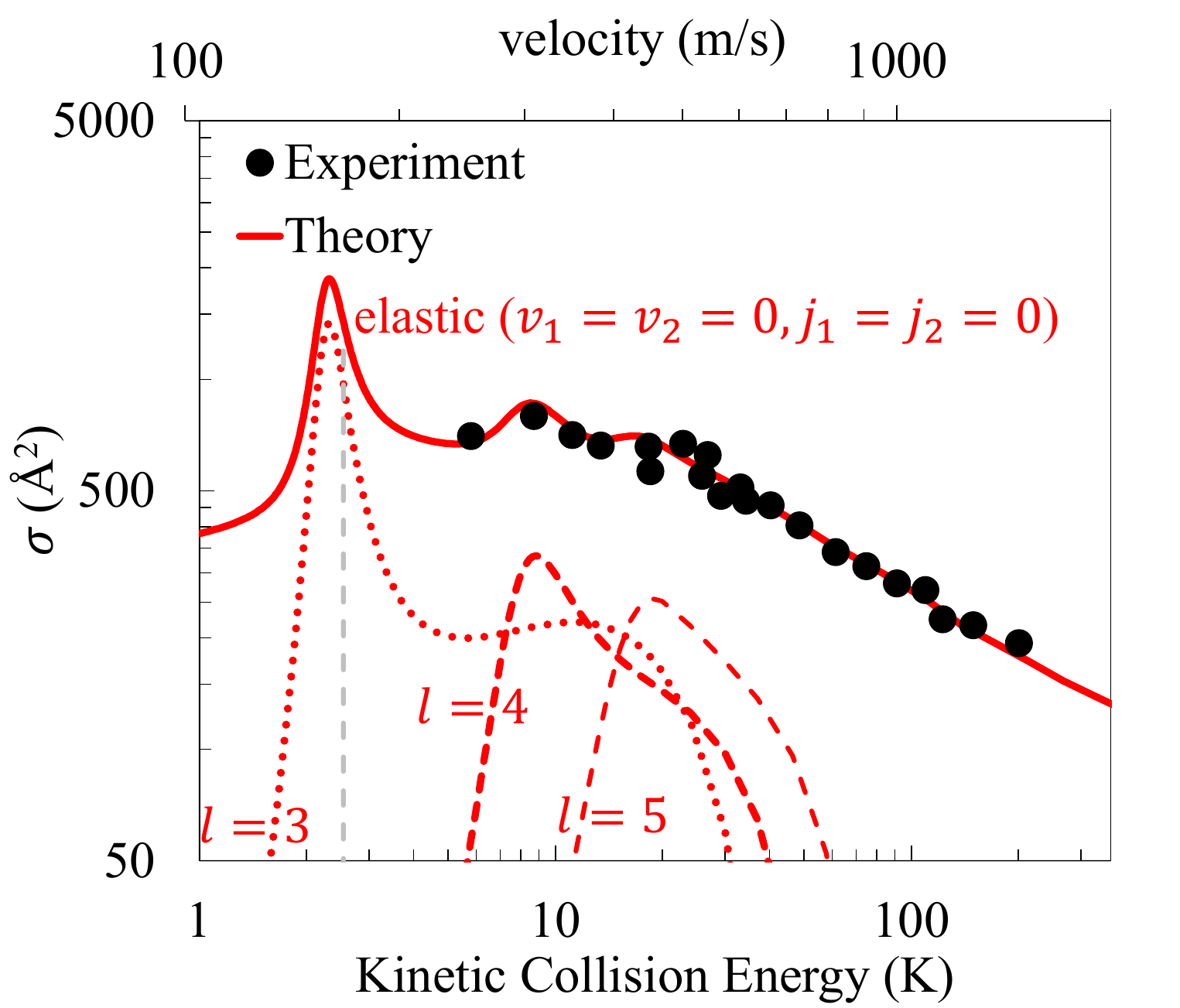}
	\caption{Integral cross sections as a function of the relative velocity from our calculations  (red curves) and experimental data of Johnson et al. (black dots and circles). Experimental results~\cite{johnson1979total} are scaled by a factor of 1.8 to enable the comparison. A partial-wave analysis of the resonance features in the cross section is also presented. The dashed gray line is the velocity corresponding to the energy of the bound state supported by the $l=3$ effective potential computed by the LEVEL code~\cite{le2017level}.}
	\label{fig:Fig_comp_exp_0000}
\end{figure}

 Recently, Cygan et al.~\cite{cygan2025dispersive} reported experimental measurements and theoretical simulations of line shape parameters for self-perturbed HD. Other experimental studies include measurements of differential cross sections for rotational excitation in collisions of ground state HD molecules~\cite{gentry1977state} and rotational energy transfer rates for vibrationally excited HD in the $v = 1$ level in collisions with thermal HD~\cite{chandler1986measurement, brown1990calculation}. However, to our knowledge, the only  experimental study of HD+HD collisions at collision energies below 100 K was by Johnson et al.~\cite{johnson1979total} nearly 40 years ago. The results are reported for the total cross sections as a function of the relative velocity between the two molecules. We assumed these cross sections correspond to the initial and final state $v_{1} j_{1} v_{2} j_{2}=v'_{1} j'_{1} v'_{2} j'_{2}=0000$ (elastic scattering  of two ground state HD molecules). We note that the experimental  values are not absolute cross sections, and thus are scaled appropriately to compare with the calculated values. Figure~\ref{fig:Fig_comp_exp_0000} displays the comparison between the experimental results of Johnson et al.~\cite{johnson1979total} and our theoretical calculations. The experimental results are scaled by a factor of 1.8  to enable the comparison. The agreement between the experiment and theory is excellent. Theory results show a peak near 150 m/s ($\sim$ 2.0 K) but the experimental data do not extend to these collision energies. A partial-wave analysis of the cross sections shows that this peak corresponds to an $l=3$ shape resonance. A strong $l=3$ resonance peak near 150 m/s was also prominently visible in the theoretical results presented by Johnson et al.~\cite{johnson1979total}, in agreement with the present results. A small resonance-like feature in both the experimental and theoretical results at about 300-400 m/s arises from the $J=l=4$ partial wave. Similarly, a minor resonance-like bump between 400-500 m/s arises from $J=l=5$. The partial-wave resolved cross sections are shown by dashed curves in Figure~\ref{fig:Fig_comp_exp_0000}.

\begin{figure}
	\centering
	\includegraphics[width=0.45\textwidth, keepaspectratio,]{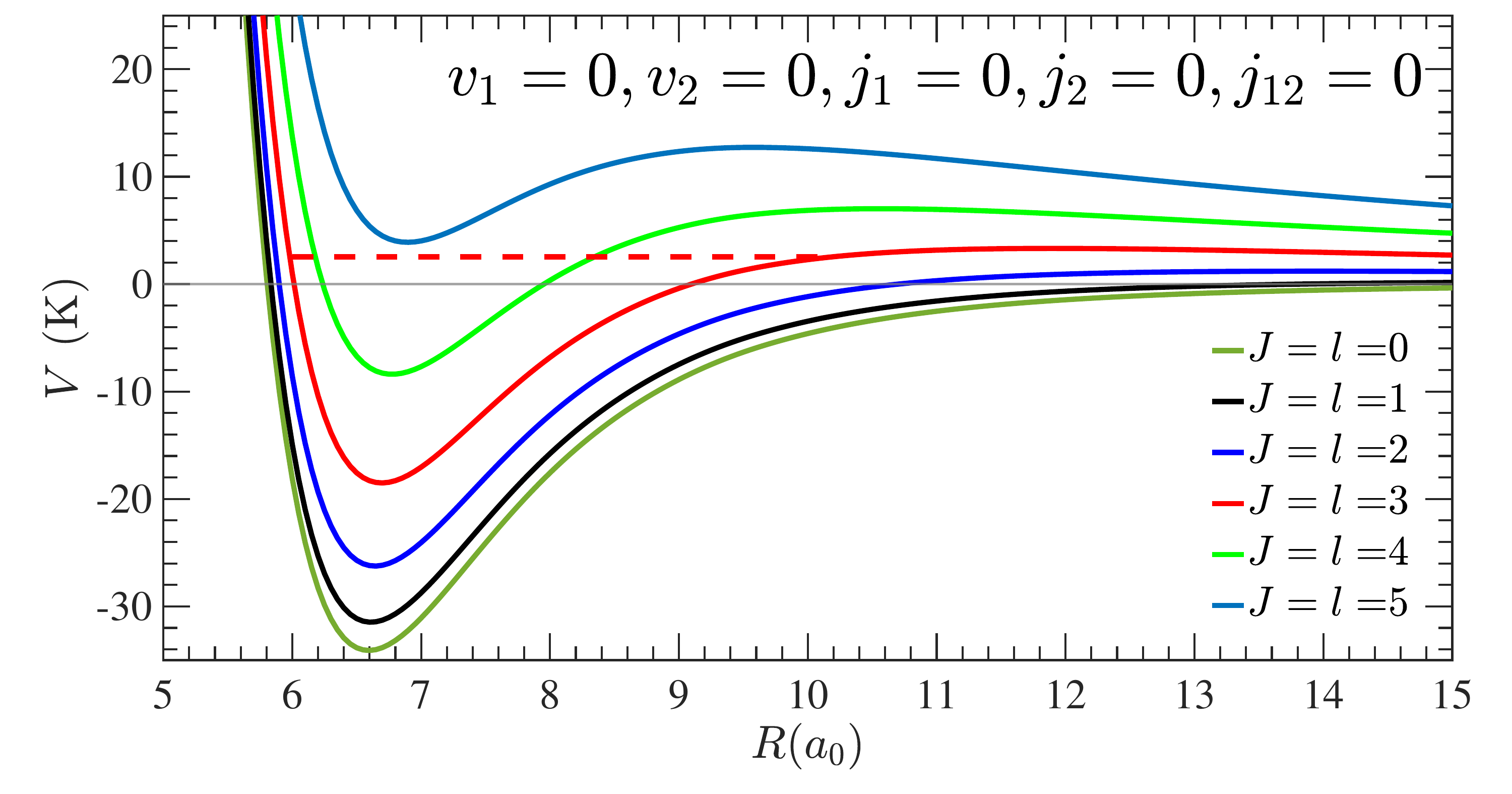}
	\caption{Diagonal terms of the effective coupling potential as a function of the intermolecular separation $R$ for the $v_{1} j_{1} v_{2} j_{2}=0000$ initial state. The different curves represent different partial waves, $l=0-5$. The dashed red line displays the energy of the $l=3$ resonance at 2.54 K computed using the LEVEL code~\cite{le2017level}.}
	\label{fig:diabats_0000}
\end{figure}

We have also examined the diagonal terms of the effective potentials (see eq. (4) of Croft et al.~\cite{Croft2018}) correlating with the initial rovibrational state  $v_{1} j_{1} v_{2} j_{2}=0000$, for $J=0-5$. This corresponds to $l=0-5$ as $j_{1}=j_{2}=j_{12}=0$. These potential curves are shown in Figure~\ref{fig:diabats_0000} as functions of $R$. While the effective potentials are similar compared to that of Johnson et al.~\cite{johnson1979total}, our theoretical cross-sections differ by roughly a factor of two (and we find no explanation for it), which necessitates the renormalization of the data from Ref.~\cite{johnson1979total}.

A bound state analysis of the $J=l=3$ potential using the LEVEL code~\cite{le2017level} identified a quasibound level at $E=2.54$ K, close to the energy of the $l=3$ resonance at $E_{c}=2.31$ K (near 150 m/s) in Figure \ref{fig:Fig_comp_exp_0000}. The energy of this resonance is indicated by the dashed red line in Fig.~\ref{fig:diabats_0000}. A similar bound state analysis of the $J=l=4$ and $j=l=5$ potential curves did not yield quasibound levels corresponding to the resonance-like features in  Figure \ref{fig:Fig_comp_exp_0000}. These features appear to be arising from the onset of these partial wave contributions.

\subsection{Inelastic rovibrational transitions}
\label{sec:Inelastic}
Our main motivation for this work is to provide more insights into  HD+HD collisions through full-dimensional quantum calculations and identify any near-resonant rotation-vibration energy transfer processes that become dominant at low collision energies.
We looked into several near resonant transitions which can occur though an exchange of rotational quantum numbers or vibrational quantum numbers of the two molecules. Namely, we looked into the following transitions for $v_{1} j_{1} v_{2} j_{2}\to v_{1} j_{2} v_{2} j_{1}$ or $v_{1} j_{1} v_{2} j_{2}\to v_{2} j_{1} v_{1} j_{2}$ considering identical particle collisions:

\begin{equation}\label{eqn:near_resonant_transitions}
	\begin{split}
		0110 \to 0011 (\Delta j= \Delta v=\pm1), \\
		0120 \to 0021 (\Delta j=\pm1,\Delta v=\pm2), \\
		0210 \to 0012 (\Delta j=\pm2,\Delta v=\pm1), \\
		0220 \to 0022 (\Delta j= \Delta v=\pm2).
	\end{split}
\end{equation}
These transitions are characterized by a small internal energy gap, primarily arising from the slightly different centrifugal distortion of the rotational levels in different vibrational levels. They also conserve the total rotational angular momentum of the two molecules, i.e., $j_{12}=j_{12}'$. The transition $0110 \to 0011$ can occur by an exchange of rotational quantum or vibrational quantum between the two molecules. The former may be referred to as quasi-resonant rotation-rotation (QRRR) transfer and the latter quasi-resonant vibration-vibration (QRVV) transfer. Similarly, the transition $0120 \to 0021$ can occur via $\Delta j=\pm1$ rotational transfer or through $\Delta v=\pm 2$ vibrational quanta exchange, while $0210 \to 0012$ can be driven by either $\Delta j=\pm 2$ or $\Delta v=\pm 1$. Lastly, for $0220 \to 0022$,  can occur through either $\Delta j=\pm 2$ or $\Delta v=\pm 2$. Because the angular anisotropy of the potential is stronger than the anisotropy with respect to the stretching of the HD bond, the primary mechanism is QRRR as discussed in our previous studies of H$_2$-H$_2$ collisions~\cite{Goulven2008,Bala_JCP_H2H2_2011}. Considering that both $\Delta j=\pm 1$ and $\Delta j=\pm 2$ transitions are allowed in HD+HD collisions one can correlate the relative efficiency of these transitions to the different anisotropic terms of the interaction potential depicted in Figure \ref{fig:Fig_exp_coeffs}. For all of these initial states, $v_{1} j_{1} v_{2} j_{2}=0110, ~0120,~0210$ and $0220$, we shall examine the relative efficiency of these near-resonant transitions against pure rotational transitions to available final channels $v'_{1} j'_{1} v'_{2} j'_{2}$:

\begin{equation}\label{eqn:pure_rotational_transitions}
	\begin{split}
		0110 \to 0010 \\
		0120 \to 0020 \\
		0210 \to 0110 \\
		0210 \to 0011 \\
		0210 \to 0111 \\
		0220 \to 0120 \\
		0220 \to 0121
	\end{split}
\end{equation}

\begin{figure}
	\centering
	\includegraphics[width=0.5\textwidth, keepaspectratio,]{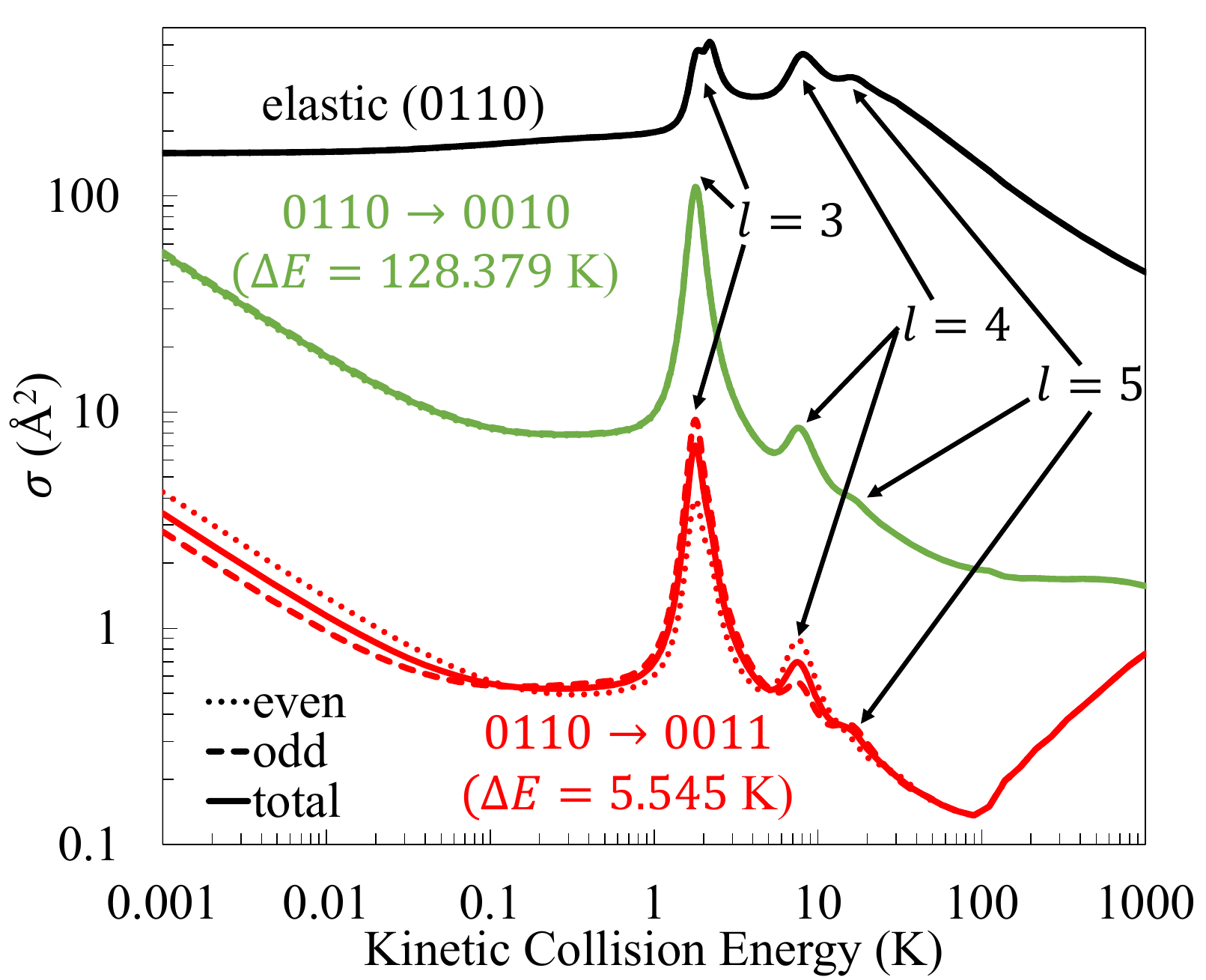}
	\caption{Integral cross sections for elastic and inelastic rovibrational transitions from initial state $v_{1} j_{1} v_{2} j_{2}=0110$ as functions of the collision energy. Different transitions are labeled by the text inside the figure along with the energy gap. Contributions from the even and odd exchange symmetries are indicated by dotted and dashed curves while their weighted sum is depicted by the solid curves for each transitions.}
	\label{fig:ini_0110}
\end{figure}

Figure~\ref{fig:ini_0110} displays the cross sections for the elastic and leading inelastic  rovibrational transitions for the initial state $v_{1} j_{1} v_{2} j_{2}=0110$ as  functions of the collision energy. The elastic cross section dominates its inelastic counterparts at all energies. The two inelastic transitions correspond to  $0110 \to 0011$ (QRRR/QRVV) and $0110 \to 0010$ (pure rotational quenching). Contributions from both even and odd exchange parities to these transitions are displayed by the dotted and dashed curves. The solid curve corresponds to the  weighted sum of these two parities considering nuclear spins of the two HD molecules~\cite{johnson1979total}:

\begin{equation}\label{eqn:HD_spin_weights}
	\sigma_{\mathrm{total}}= \frac{15}{36}\sigma_{\mathrm{even}} + \frac{21}{36}\sigma_{\mathrm{odd}}.
\end{equation}

The near-resonant transition, $0110 \to 0011$ has a tiny energy gap of $\Delta E\simeq5.5$ K compared to $\Delta E\simeq128$ K for the pure rotational transition, $0110 \to 0010$ as specified  in  Fig~\ref{fig:ini_0110}. Despite the smaller energy gap for the $0110 \to 0011$ QRRR/QRVV transition its cross section  is  an order of magnitude smaller compared to the $0110 \to 0010$ transition. This can be correlated to the strength of the expansion coefficients $C_{\lambda_{1}, \lambda_{2}, \lambda_{12}}$ (Figure~\ref{fig:Fig_exp_coeffs}) corresponding to $\lambda_{1}, \lambda_{2}$, $\lambda_{12}=0,1,1$ that drives the pure rotational transition. Its magnitude is larger  compared to $\lambda_{1}, \lambda_{2}$, $\lambda_{12}=1,1,0$ or $1,1,2$ terms responsible for the near-resonant transition. Thus, in this case, the anisotropy of the potential becomes a deciding factor in driving rotational transitions.

A partial-wave analysis of different peaks in Fig.~\ref{fig:ini_0110} shows that the primary resonance peak corresponds to $l=3$ near $E_{c}\simeq2$ K. The secondary peak due to  $l=4$ occurs around $E_{c}\simeq8$ K. For the elastic transition, the $l=3$ and $l=4$ peaks are of comparable magnitude. A minor feature corresponding to $l=5$  is barely visible around  $E_{c}\simeq20$ K.

\begin{figure}
	\centering
	\includegraphics[width=0.5\textwidth, keepaspectratio,]{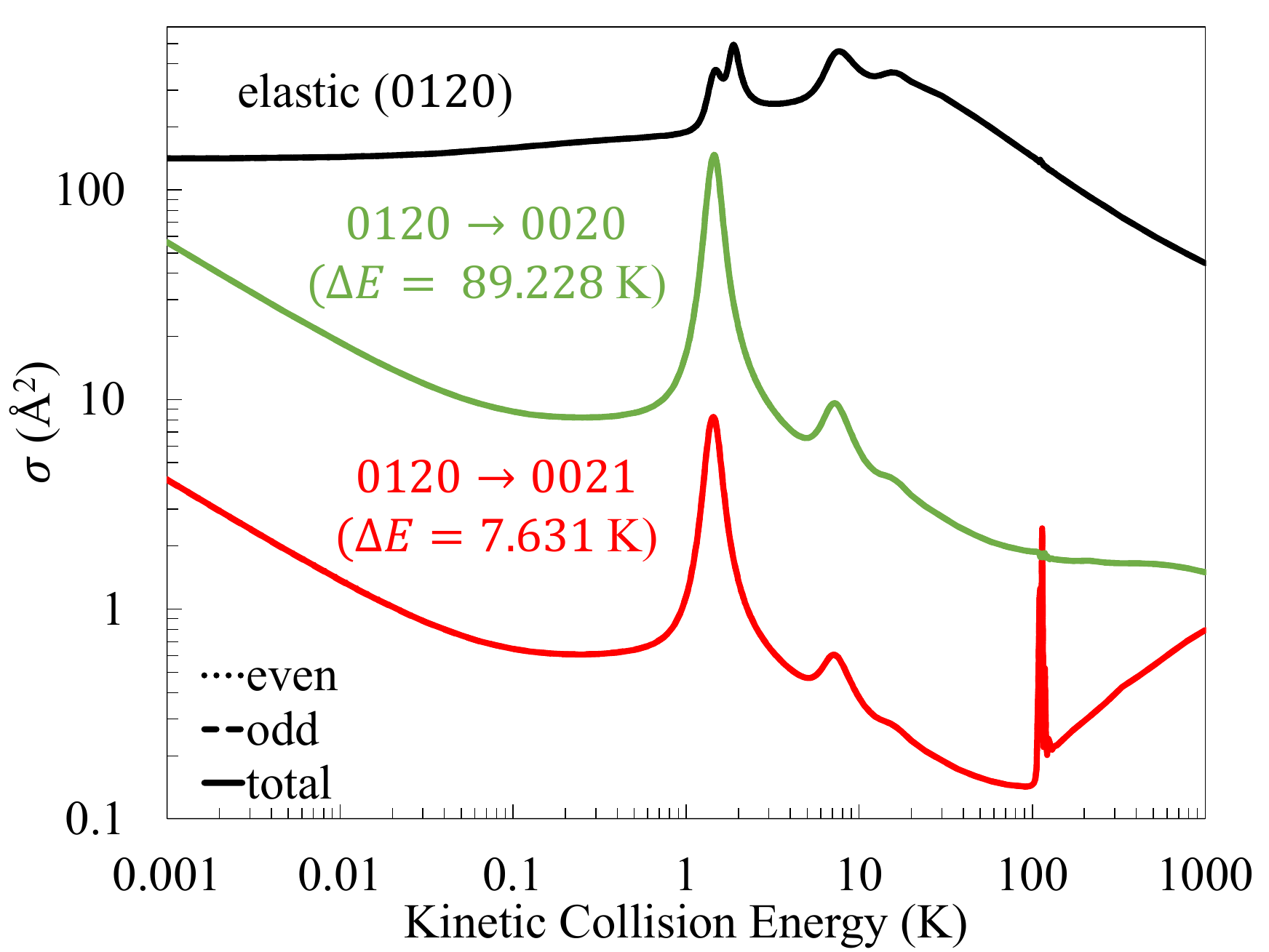}
	\caption{Same as Fig.~\ref{fig:ini_0110}, but for the initial state $v_{1} j_{1} v_{2} j_{2}=0120$. The contributions from even and odd exchange symmetries are nearly identical to the weighted sum, rendering the individual curves indistinguishable on this scale.}
	\label{fig:ini_0120}
\end{figure}

\begin{figure}
	\centering
	\includegraphics[width=0.3\textwidth, keepaspectratio,]{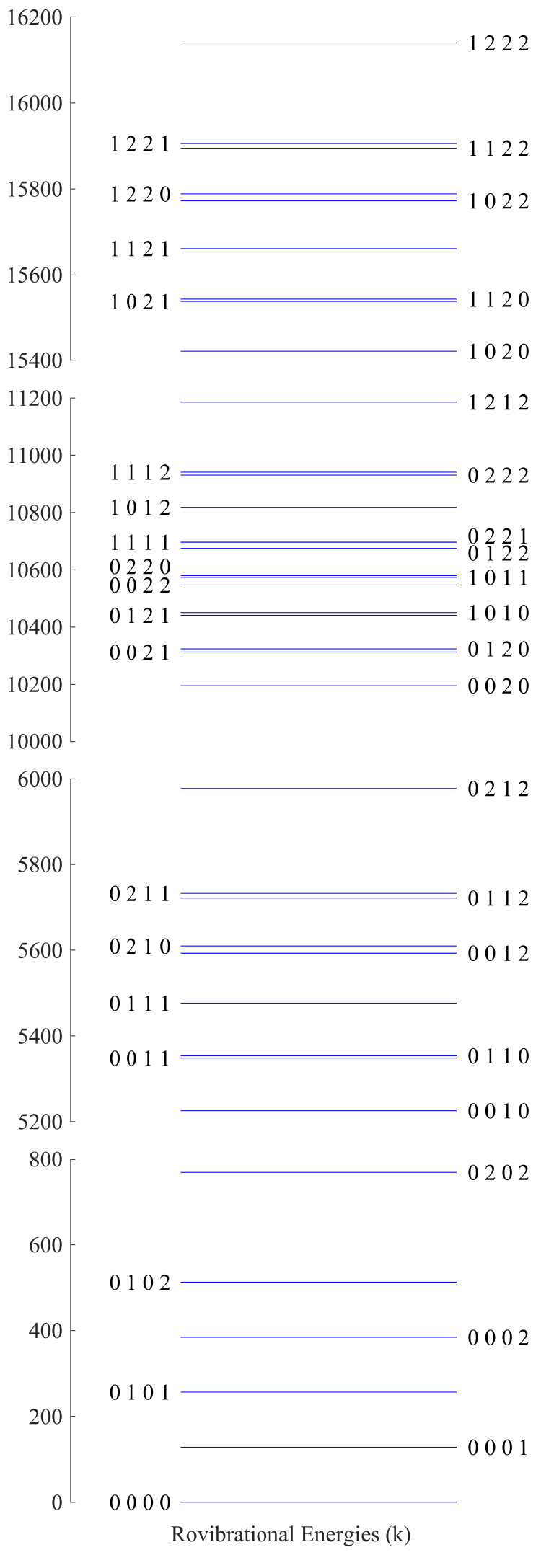}
	\caption{Energies of rovibrational states of HD+HD CMSs. All the energies are shifted with respect to the initial rovibrational level $v_{1} j_{1} v_{2} j_{2}=0000$, where energy of state 0000 is $E_{0000}=0$ K and the vertical axis represent the relative energies in kelvin.}
	\label{fig:Energy_Levels}
\end{figure}

Results of elastic and inelastic cross sections for the initial state $v_{1} j_{1} v_{2} j_{2}=0120$ presented in Fig.~\ref{fig:ini_0120} display similar features as the $0110$ state in Fig.~\ref{fig:ini_0110}. A partial wave analysis revealed the same orbital angular momentum as in $0110$ contributing to the resonance peaks. A notable feature in this case is a sharp peak near a collision energy of $E_{c}=114$ K. This appears to be  a Feshbach resonance due to a quasibound state supported by the 0121 channel which opens at an energy of 117 K with respect to the asymptotic energy of the 0120 channel. A schematic representation of the rovibrational energy levels  of the CMSs of HD+HD is presented in Figure~\ref{fig:Energy_Levels}. These energies are relative to the energy of the  $v_{1} j_{1} v_{2} j_{2}=0000$ CMS which is taken to be $E_{0000}=0$ K. The energy gap between the 0120 and 0121 channels is nearly 117 K as Fig.~\ref{fig:Energy_Levels} illustrates.

To confirm that this is indeed a Feshbach resonance, we removed the 0121 state from  basis set, and thus,  coupling to this channel at short-range. This eliminated the resonance peak in the  $0120 \to 0021$ cross section and also the tiny feature at the same energy in the elastic cross section. We believe this feature occurs prominently for the $0120 \to 0021$ transition because its background cross section is small compared to the elastic and the  $0120 \to 0020$ transition. A similar feature is also observed in the cross section for the  $0220 \to 0121$ transition, as discussed below. Similar Feshbach resonances were previously reported for rotationally inelastic collisions in CO+He~\cite{balakrishnan2000vibrational}.

\begin{figure}
	\centering
	\includegraphics[width=0.5\textwidth, keepaspectratio,]{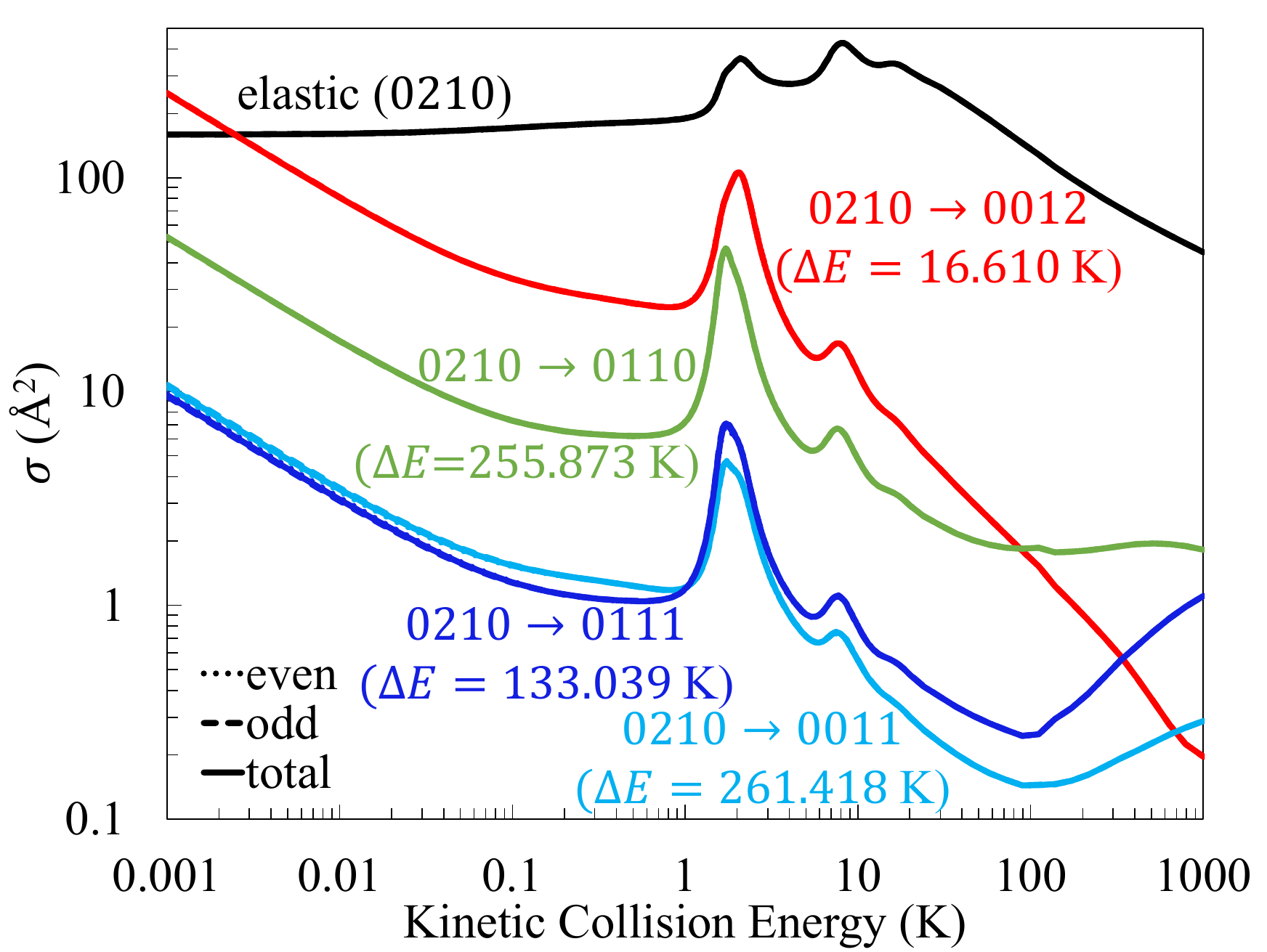}
	\caption{Same as Fig.~\ref{fig:ini_0120}, but for the initial state $v_{1} j_{1} v_{2} j_{2}=0210$. The legends inside the figure indicate the rovibrational transitions and the corresponding energy gaps. The dotted and dashed curves of colors correspond to the even and odd exchange symmetries, respectively, while solid curves show the weighted sum. However, the contributions from even and odd exchange symmetries are nearly identical to the weighted sum, rendering the individual curves indistinguishable on this scale.}
	\label{fig:ini_0210}
\end{figure}

 Figure~\ref{fig:ini_0210} displays the energy dependent cross sections for elastic and inelastic transitions from the initial state $v_{1} j_{1} v_{2} j_{2}=0210$. As given by eqs. (\ref{eqn:near_resonant_transitions} and \ref{eqn:pure_rotational_transitions}), the near-resonant transition is $0210 \to 0012$, while several pure rotational  transitions involving quenching and/or excitations are possible, namely $0210 \to 0110$, $0210 \to 0011$, and $0210 \to 0111$. In this case, the QRRR  transfer is the dominant inelastic transition compared to the other rotationally inelastic transitions for energies below 100 K.

The dominance of the near-resonant transition in this case can also be ascribed to the magnitude of the anisotropic term that drives the transition. In this case, the process involves exchange of two rotational quanta ($\Delta j =2$) and it is primarily driven the  term $\lambda_{1}, \lambda_{2}$, $\lambda_{12}=2,2,4$, which is larger than all other anisotropic terms in the angular dependence of the interaction potential (see Figure~\ref{fig:Fig_exp_coeffs}). This combined with its smallest energy gap  ($\Delta E\simeq17$ K) compared to other rotational transitions ($\Delta E\simeq256$ K for the next largest inelastic transition leading to the 0110 final state) make it the dominant inelastic transition. While the $0210 \to 0111$ is also a QRRR transition involving $\Delta j=\pm 1$ its magnitude is small due to the smaller anisotropic term that drives this transition as discussed for the $0110 \to 0011$ transition.

A partial wave analysis reveals that the primary peak in the inelastic cross sections near $E_{c}\simeq2$ K arises from $l=3$, while an $l=4$ peak appears around $E_{c}\simeq8$ K. A  minor bump corresponding to $l=5$  is barely visible around $E_{c}\simeq20$ K. For the elastic transition, the $l=4$ resonance peak is slightly larger in magnitude compared to $l=3$ peak.

\begin{figure}
	\centering
	\includegraphics[width=0.5\textwidth, keepaspectratio,]{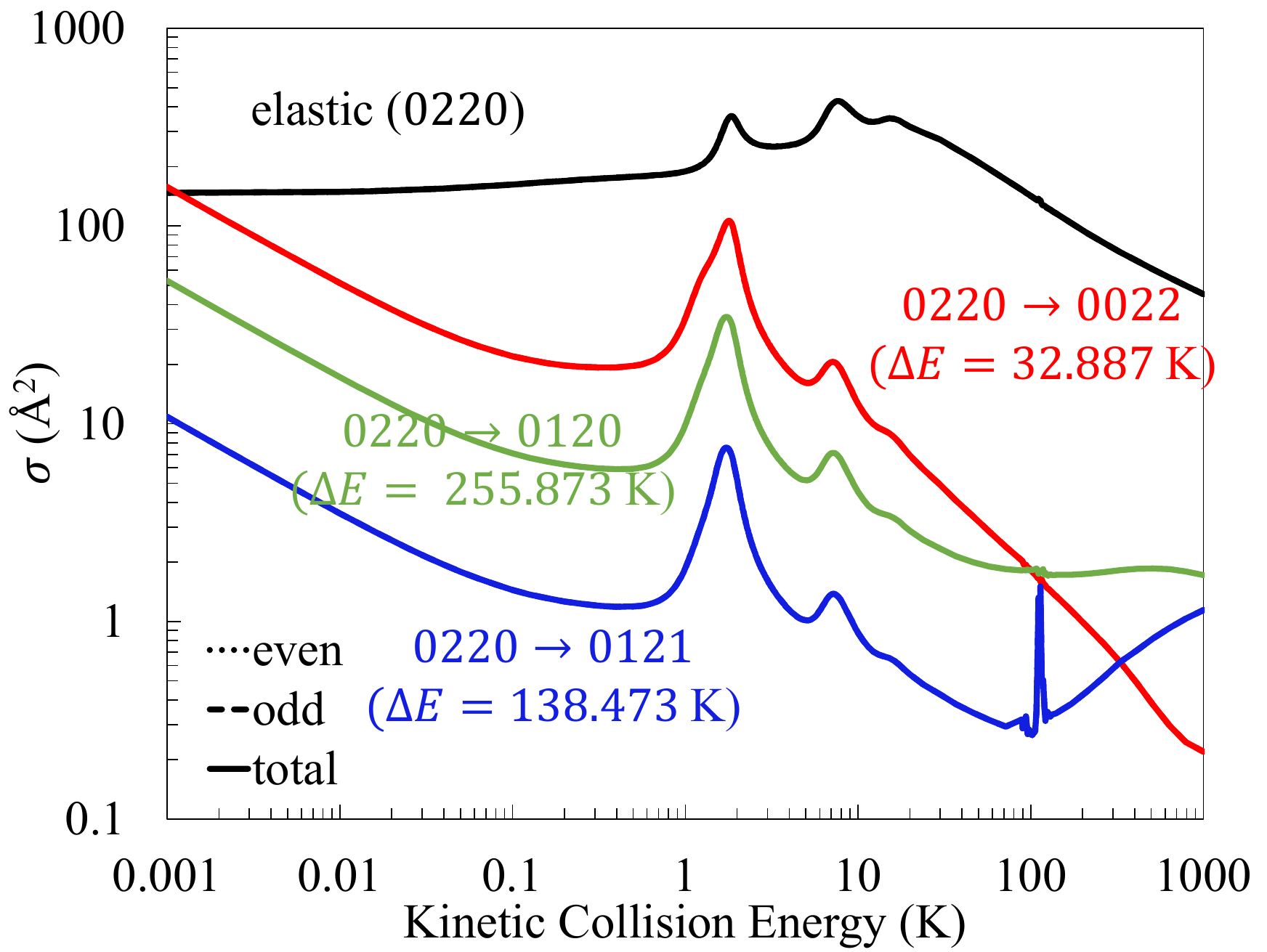}
	\caption{Same as Fig.~\ref{fig:ini_0210}, but for the initial state $v_{1} j_{1} v_{2} j_{2}=0220$. The contributions from even and odd exchange symmetries are nearly identical to the weighted sum, rendering the individual curves indistinguishable on this scale.}
	\label{fig:ini_0220}
\end{figure}

For the initial state $v_{1} j_{1} v_{2} j_{2}=0220$ the near-resonant transition leading to the 0022 final state can occur through QRRR process involving $\Delta j=2$ or a QRVV process involving $\Delta v=2$ though the former process is expected to dominate for reasons explained earlier in this section. Figure~\ref{fig:ini_0220} displays the ICS for the various inelastic transitions given by eqs. (\ref{eqn:near_resonant_transitions}-\ref{eqn:pure_rotational_transitions}) from this state. The near-resonant transition $0220 \to 0022$ is  about 4 times larger than the pure rotational  transition $0220 \to 0120$ and over an order of magnitude greater than the final state 0121 (also a QRRR transition) for energies lower than 10 K. The mechanism behind this is the same as explained before - the dominance of the $\lambda_{1}, \lambda_{2}$, $\lambda_{12}=2,2,4$ term that drives this transition combined with its smaller energy gap. The partial-wave contributions to the resonance peaks are similar to that for the initial state 0210 depicted in Figure~\ref{fig:ini_0210} and is not elaborated further.

A Feshbach resonance in the $0220 \to 0121$ cross section near $E_{c}\simeq114$ K appears due to the opening of the 0221 channel at an energy of 117 K relative to the threshold of the 0220 state. This is similar to the Feshbach resonance discussed earlier in the cross section for the $0120 \to 0021$ transition. As before, an analysis that omits the 0221 channel in the basis set  eliminated this resonance, confirming it to be a Feshbach resonance. A similar Feshbach resonance is also observed in the cross sections for the $0220 \to 0021$ transition but its magnitude is smaller than that of $0220 \to 0121$.

Similar to eq. (\ref{eqn:near_resonant_transitions}), there are a few more initial states within $v=0-2$  and $j=0-2$ for which near-resonant transitions are possible. These transitions, including those from eq. (\ref{eqn:near_resonant_transitions}), are listed in Table~\ref{table:all_near_resonant_transitions}. All of these QRRR transitions exhibit similar patterns in the collision energy dependence of their  cross sections. For QRRR  transitions that involve a $\Delta j= \pm1$ rotational transfer, the cross sections are consistently smaller  compared to pure rotational transitions, irrespective of the vibrational quantum number. For example, cross sections for $0211\to0112$, $0221\to0122$, $1221\to1122$, and $1120\to1021$ QRRR transitions are smaller in magnitude compared to $0211\to0210$ or $0211\to0111$, $0221\to0121$ or $0221\to0220$, $1221\to1220$ or $1221\to1121$, and $1120\to1020$ pure rotational transitions. This is because the expansion coefficient that drives these transitions is smaller in magnitude compared to that for pure rotational transitions as discussed in details earlier. On the other hand, the QRRR transition $1220\to1022$ that involve $\Delta j=\pm2$ rotational transfer, has a larger cross sections compared to pure rotational quenching, $1220\to1120$, due to the dominance of the $\lambda_{1}, \lambda_{2}$, $\lambda_{12}=2,2,4$ term involved in the QRRR process.

We also found that if one of the identical HD molecules is in an initial rovibrational level of $v=2,j=0$, then it exhibits a Feshbach resonance supported by a rotationally excited state of the same HD molecule, $v=2,j=1$. We discussed these Feshbach resonances for the 0120 and 0220 initial states earlier, but we also found similar results for 1120 and 1220. The energy gap between these initial states and the excited states is always about $\Delta E \simeq 117.3997$ K as indicated in Table~\ref{table:all_near_resonant_transitions}.

\begin{table}
	\caption{\label{table:all_near_resonant_transitions} List of near-resonant transitions.}
	\centering
	\resizebox{\linewidth}{!}{%
		\begin{tabular}{c c c c c}
			\hline
			\hline
			Initial state & Final state & $|\Delta j|$ & $|\Delta v|$ & State that supports a Feshbach resonance \\
			\hline
			\hline
			0110 & 0011 & 1 & 1 &None \\
			0210 & 0012 & 2 & 1 & None\\
			0211 & 0112 & 1 & 1 &None \\
			0221 & 0122 & 1 & 2 & None\\
			1221 & 1122 & 1 & 1 &None \\
			0120 & 0021 & 1 & 2 & 0121 ($\Delta E_{0121-0120}=$117.3997 K) \\
			0220 & 0022 & 2 & 2 & 0221 ($\Delta E_{0221-0220}=$117.3997 K) \\
			1120 & 1021 & 1 & 1 & 1121 ($\Delta E_{1121-1120}=$117.3997 K) \\
			1220 & 1022 & 2 & 1 & 1221 ($\Delta E_{1221-1220}=$117.3997 K) \\
		\end{tabular}%
	}
\end{table}

\begin{figure}
	\centering
	\includegraphics[width=0.5\textwidth, keepaspectratio,]{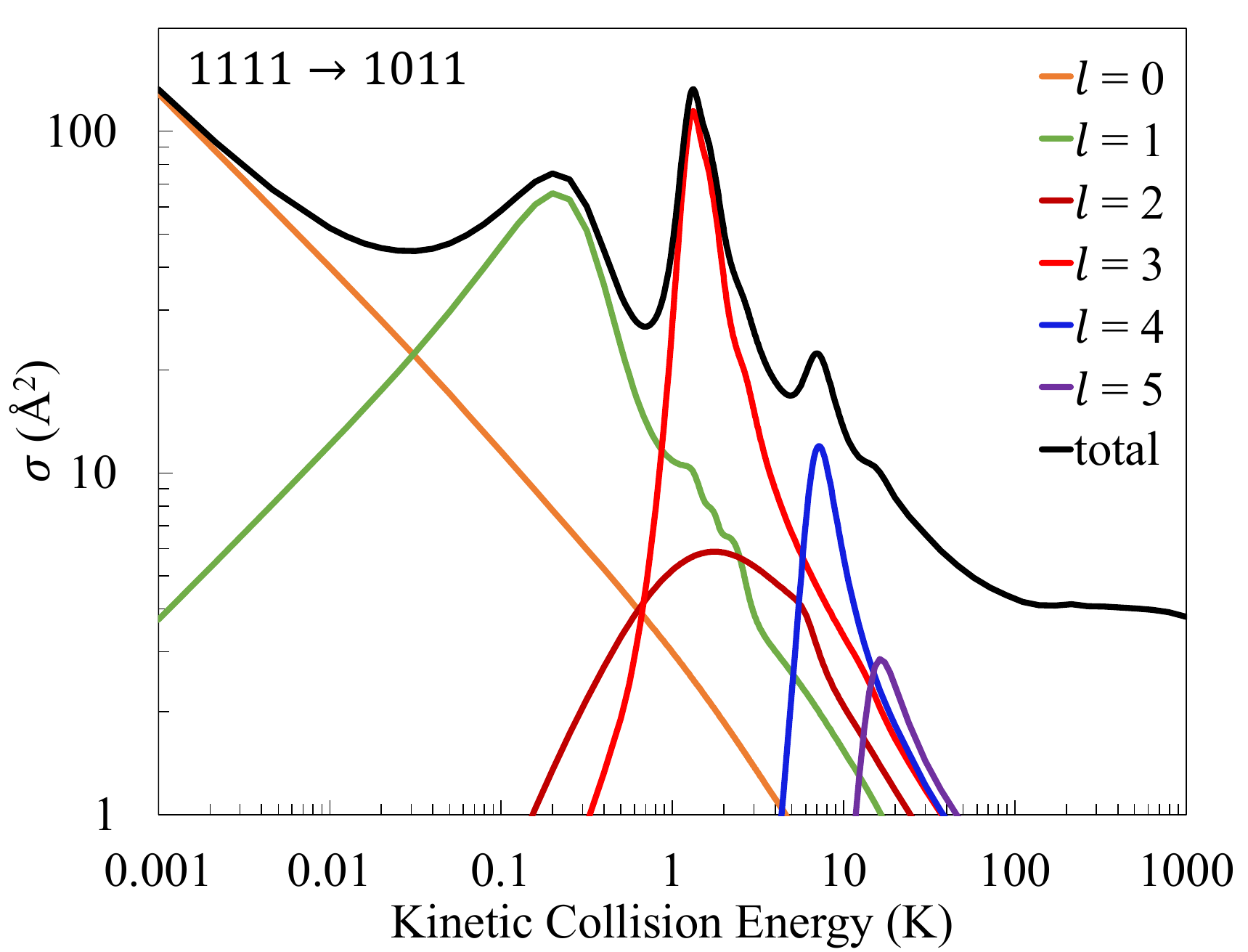}
	\caption{Partial-wave resolved cross sections for the $1111 \to 1011$ rotational transition as a function of collision energy, illustrating contributions from an $l=1$ and $l=3$ shape resonances.}
	\label{fig:L_resolved_1111_1011}
\end{figure}

Apart from these initial states, we  examined inelastic transitions from several other rotational states in different vibrational levels. Results for these initial states $v_{1} j_{1} v_{2} j_{2}=0202, 1111, 1212$ and $2222$, are presented in the Supplementary Material.  In all these cases, pure rotational transitions dominate over vibrational transitions. Unlike other initial states discussed so far, for which resonance features mostly occur for $l=3-5$ partial waves,  for  $v_{1} j_{1} v_{2} j_{2}=1111$, a partial-wave analysis  shows an $l=1$ resonance near 0.2 K followed by an $l=3$ resonance between $1-2$ K. This is shown for the  $1111 \to 1011$ transition in Figure~\ref{fig:L_resolved_1111_1011}. Compared to the $0110\to0010$ pure rotational quenching cross sections shown in Figure~\ref{fig:ini_0110}, the cross sections for the $1111 \to 1011$ transition is a factor of two larger as the s-wave limit approaches for energies below $10^{-3}$ K.

\subsection{Temperature dependent rate-coefficients}
\begin{figure}
	\centering
	\includegraphics[width=0.45\textwidth, keepaspectratio,]{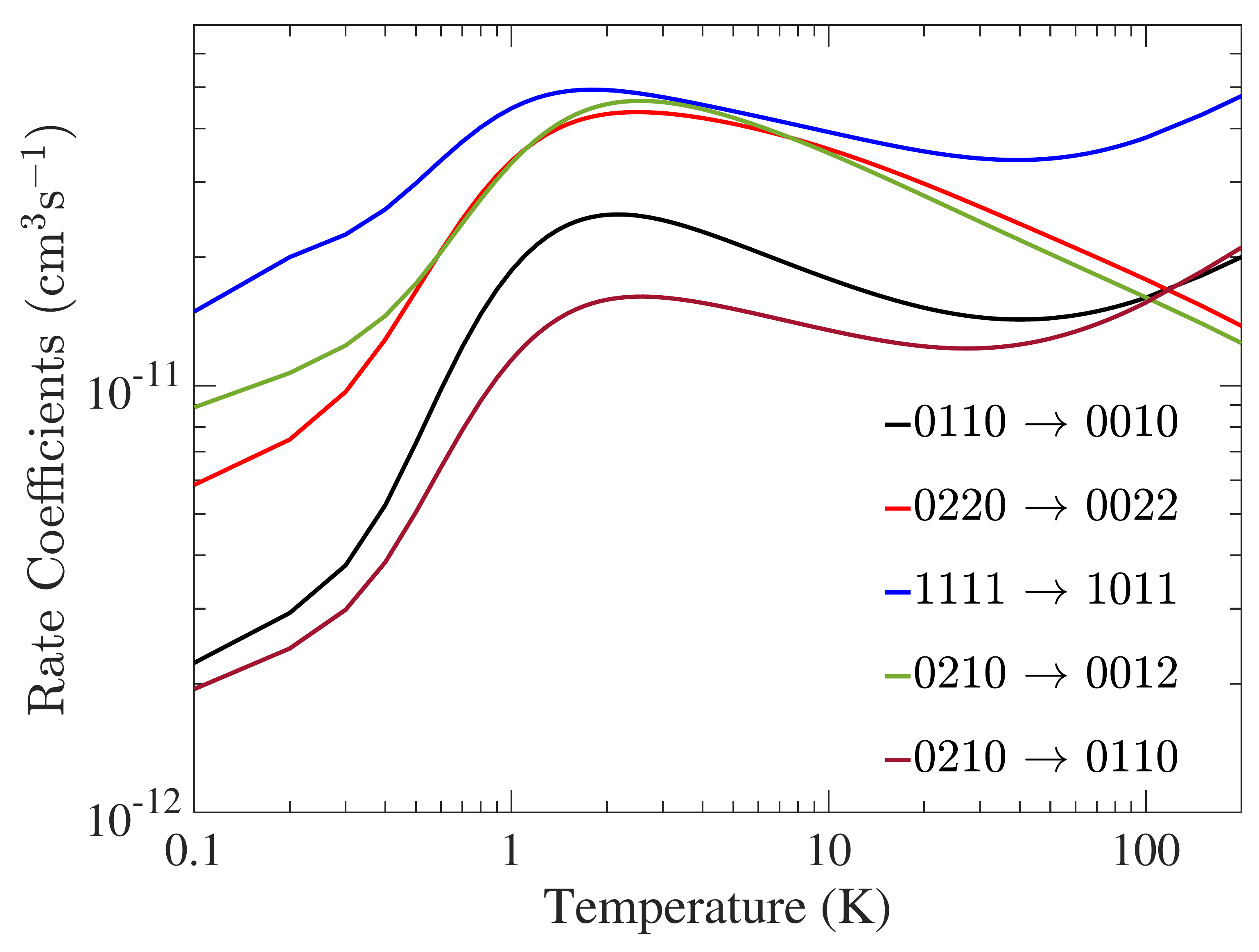}
	\caption{Temperature dependent rate-coefficients for several of the dominant inelastic transitions. The curves are labeled by their initial and final CMSs as indicated by the figure legends.}
	\label{fig:rate_coefficients_inelastic_selective}
\end{figure}

Rate-coefficients for several dominant inelastic transitions, including pure rotational as well as quasi-resonant transitions, as a function of the temperature are displayed in Figure~\ref{fig:rate_coefficients_inelastic_selective}. These rate-coefficients correspond to some of the dominant cross sections  discussed in Section \ref{sec:Inelastic}. Different transitions are indicated by different color curves as mentioned in the figure legends.

There are two subgroups among the rate coefficients shown in Fig.~\ref{fig:rate_coefficients_inelastic_selective}. The first group corresponds to  rotationally inelastic transitions, shown by the blue, brown and black curves  representing $1111\to1011$, $0210\to0110$ and $0110\to0010$, respectively. They involve one quantum rotational quenching. The rate-coefficients generally increase with temperature, reflecting the pattern of the corresponding cross sections. The other group refers to the quasi-resonant transitions where an exchange of two rotational quanta occur between molecules in the $v=0$ and $v=1$ or $v=2$ vibrational levels.  They correspond to $0210\to0012$ and $0220\to0022$  transitions, depicted by green and red curves, respectively. They also exhibit a peak in the 1-5 K range due to the $l=3$ resonance but  their magnitude decreases with further increase of temperature, illustrating the dominance of these transitions at lower collision energies. 
			
\section{Summary and Conclusions}
\label{sec:summary}
In this manuscript, we report the first full-dimensional quantum calculations of ro-vibrational transitions in HD+HD collisions on a highly accurate interaction potential for the H$_4$ system. Results for total cross sections (primarily elastic collisions) are found to be in close agreement with the experimental results of Johnson et al.~\cite{johnson1979total} reported in 1979, including a resonant peak arising from an $l=3$ partial wave predicted by their theoretical calculations.

 Inelastic collisions are dominated by pure rotational transitions and near-resonant transitions that involve exchange of two rotational quanta between the two HD molecules. The magnitude of the rotational quenching cross sections are correlated to the strength of the different anisotropic terms of the interaction potential, $C_{\lambda_{1}, \lambda_{2}, \lambda_{12}}$, presented in Figure~\ref{fig:Fig_exp_coeffs}. In particular, the near-resonant transitions are primarily driven by the  $\lambda_{1}, \lambda_{2}$, $\lambda_{12}=2, 2, 4$ terms, which have the largest magnitude among all of the anisotropic  terms. Further, these transitions are characterized by a small energy gap and conservation of internal rotational angular momentum of the two molecules. Regardless of the nature of the rotational transitions, resonant part of the cross sections is dominated by an $l=3$ partial wave. An $l=1$ partial-wave resonance is found to occur only for pure rotational transitions from the $v_{1} j_{1} v_{2} j_{2}=1111$ initial state. 	
 
 Recent experimental studies of HD+H$_2$/D$_2$ and D$_2$+D$_2$ collisions by Zare and co-workers in which one or both molecules were prepared in a specific rovibrational quantum state, and with specific alignments of their bond-axis, through the SARP techniques   have attracted much attention. While HD+HD collisions are yet to be explored by the SARP techniques, the results reported here for isotropic collisions suggest that HD+HD collisions offer a fertile ground for probing alignment effects on quasi-resonant transitions dominated by quantum resonances near 1 K. Energy-resolved SARP experiments can offer much insights into the leading anisotropic terms that drive these transitions. We hope such experiments would be pursued in the near future.
 
\section*{Supplementary Material}
The Supplementary Material includes collision energy dependent cross sections for  initial states, $v_{1}j_{1}v_{2}j_{2}=0202,1111,1212,$ and $2222$ of the two HD molecules, depicting both pure rotational and ro-vibrational transitions.
	
\begin{acknowledgments}
	This work is supported in part by  grant No. PHY-2409497 (N. B.) from NSF and the European Union (Grant No 101075678, ERC-2022-STG, H2TRAP). Views and opinions expressed are, however, those of the author(s) only, and do not necessarily reflect those of the European Union or the European Research Council Executive Agency. Neither the European Union nor the granting authority can be held responsible for them.
\end{acknowledgments}
	
\section*{Conflicts of interest}
There are no conflicts to declare.

		
\section*{Data availability statement}
The data that support the findings of this study are available within the article and its Supplementary Information.

\end{document}